\documentclass{article}\usepackage[utf8]{inputenc}
\usepackage{float}
\usepackage{amsmath}
\usepackage{amssymb}
\usepackage{esvect}
\usepackage[letterpaper, margin=0.7 in]{geometry}
\usepackage{authblk}
  \usepackage{amsmath}
    \usepackage{amssymb}
    \usepackage{amsthm}
    \usepackage{amsfonts}
    \usepackage{braket}
\usepackage{amsmath}
\usepackage{amsfonts}
\usepackage{amssymb}
\usepackage{hyperref}
\usepackage{graphicx}
\usepackage{caption}
\usepackage{subcaption}
\usepackage{float}
\usepackage{mathtools}

\title{Repulsive Casimir force in Bose-Einstein Condensate\\}
\author{Mir Mehedi Faruk$^{1}$\thanks{Corresponding author: muturza3.1416@gmail.com,  mir.faruk@mail.mcgill.ca},
} 
\author{Shovon Biswas$^{2}$}

\affil{
McGill University
Montreal, QC H3A 2T8, Canada$^1$\\
 Department of Electrical and Electronic Engineering, Bangladesh University of Engineering and Technology, Dhaka
1000, Bangladesh.$^2$
}

\begin{document}
\maketitle
\begin{abstract} 
We study the  Casimir effect for a three dimensional system of ideal free massive Bose gas in a slab
geometry with Zaremba and anti-periodic boundary conditions.
It is found that for these type of boundary conditions the resulting Casimir force is repulsive in nature,  in contrast with usual periodic, Dirichlet or Neumann boundary condition where the Casimir force is attractive (Martin P. A. and Zagrebnov V. A., Europhys. Lett.,
73 (2006) 15.).
Casimir forces in these boundary conditions also maintain a
power  law decay  function below condensation temperature and exponential 
decay function above the condensation temperature albeit with a positive sign, identifying the repulsive nature of the force.\\

\end{abstract}
\section{Introduction}
In his original paper, Casimir 
described  a nonclassical
attractive force  related to  quantum vacuum fluctuations in the electromagnetic field between two uncharged
parallel conducting plates\cite{casimir}.
Since then, the
Casimir effect for quantum vacuum fluctuation has been extensively studied for various types of geometries and boundary conditions (see ref. \cite{2}, \cite{3} and the references therein) using Quantum Field Theory (QFT) techniques.
But the  Casimir type force due to \textit{thermal fluctuation} in ideal free Bose gas in vacuum was first  reported in the seminal work of Martin and Zagrebnov\cite{4} and since then it has been thoroughly studied\cite{5,6,marek1,marek2,marek4,marek5} in Statistical Mechanics (SM). It is quite well known that the Casimir force  depends upon boundary conditions and is attractive
for scalar fields (as well as free Bose gas) in either case of vacuum or thermal fluctuation and is
reported to be  an attractive force for 
the usual case of Dirichlet (D),
Neumann (N) as well as periodic boundary (P) 
conditions on both sides.
 However, it is of significant interest to get physical configurations
where the Casimir force is repulsive instead of attractive, not only for its relevance for technical
applications to nano devices\cite{3,nano,milton33,app,app2}, but also because the existence of repulsive
or null Casimir forces allows a more accurate analysis of micro-gravity effects\cite{app3}
as well as the study of cosmic strings\cite{string}. It has been recently reported  that in QFT approach one can achieve repulsive Casimir force due to quantum fluctuation using Zaremba\footnote{Zaremba boundary condition indicates Dirichlet condition in one side and Neumann condition in another side, also known as hybrid or mixed boundary condition.} and anti-periodic boundary condition\cite{qft1,qft2} (see section 3.4 and 3.5 of ref. \cite{qft2}). But  Casimir type force due to thermal fluctuation in free Bose gas with these type of boundary condition is yet to be reported.
 Casimir-type interactions are
nowadays identified in several types  of systems spanning from biology\cite{bio}  to cosmology\cite{cos} but the
QED and condensed-matter contexts are those where
the theoretical predictions concerning the existence and
properties of Casimir forces found firm experimental confirmation\cite{experiment,ex2}. Therefore, it is of significant importance to figure out if Zaremba and anti-periodic boundary conditions can be responsible  for repulsive Casimir force in a free Bose gas due to thermal fluctuations in SM.
In this manuscript we have considered  an ideal free
Bose gas confined in 
three dimensional slab like geometry $L\times L \times d$ (where $L>>d$), subjected to  Zaremba/anti-periodic boundary condition in $z$ direction and calculated the repulsive Casimir force due to thermal fluctuation in vacuum. 
Point to note,   \textit{imperfect}  Bose gas 
with 
the repulsive
microscopic interparticle interactions
subjected to  periodic boundary conditions generate 
an effective Casimir force of repulsive nature\cite{marek5}. But in that case\cite{marek5}, the repulsive behaviour of Casimir force
is due to the  interparticle interaction whereas in the present endeavor
the repulsive Casimir force is solely due to boundary condition.
Relation between the decay length
characterizing 
the Casimir force
(due to these boundary condition)
and the bulk correlation lengths 
are also discussed.\\

\section{Model}
Let us consider ideal free massive Bose gas  confined between two infinitely large square shaped  plates
of area $A$. The plates are along the
$xy$ plane and are separated by  
distance $d$
along the z-axis. For the slab geometry
we consider 
$\sqrt{A}>>d$
as well the 
system is in thermodynamic
equilibrium with its surroundings at temperature $T$. 
At this temperature the thermal de Broglie
wavelength of a single particle of mass $m$ is  $\lambda=\hbar\sqrt{\beta/m}$ where $\beta=\frac{1}{k_BT}$ and $k_B$ is the Boltzmann's constant. In the thermodynamic limit, we consider
$\lambda<<d$. Thus the energy of the single particle is $E=\frac{ q_x^2}{2m}+\frac{q_y^2}{2m}+\frac{ p_z^2}{2m}$, where for Zaremba (Z) and antiperiodic (A)  boundary condition\cite{qft1,qft2} we have respectively,
\begin{subequations}
\begin{eqnarray}
  &&  p_z=\left(n+\frac{1}{2}\right)\frac{\hslash \pi}{d}, \quad
n=0,1,2,3,...\\
 &&p_z=\left(n+\frac{1}{2}\right)\frac{2\pi\hslash}{d}, \quad
n=0,\pm1,\pm2,\pm3,...
\end{eqnarray}
\end{subequations}
Based on the assumptions described above the grand-canonical potential per unit area can be written as
\begin{eqnarray}
    \Phi_d(T,\mu) =\beta^{-1}
    \sum_{n=0}^{\infty}\int_{-\infty}^{\infty}\int_{-\infty}^{\infty}\frac{dq_x dq_y}{(2\pi\hbar)^2}
   \Omega(q_x,q_y,p_z)
\end{eqnarray}
where,
\begin{align}
\Omega = \left\{ \begin{array}{cc} 
                 \ln\left[1-e^{-\beta\left(\frac{q_x^2}{2m}+\frac{q_y^2}{2m}+\frac{\pi^2\hbar^2 \left(n+\frac{1}{2}\right)^2}{2md^2}-\mu\right)}\right] & \hspace{5mm} \text{(Z)} \\
                \ln\left[1-e^{-\beta\left(\frac{q_x^2}{2m}+\frac{q_y^2}{2m}+\frac{2\pi^2\hbar^2 \left(n+\frac{1}{2}\right)^2}{md^2}-\mu\right)}\right] & \hspace{5mm} \text{(A)} \\
              \end{array} \right.
\end{align}

Here $\mu$ is the chemical potential.
Representing (2) by its low-activity series for $\mu<0$ and performing the integration we obtain,

\begin{align}
\Phi_d(T,\mu) = \left\{ \begin{array}{cc} 
                 -\frac{1}{2\pi \beta\lambda^2}\sum\limits_{r=1}^{\infty}\frac{e^{\beta r \mu}}{r^2}\sum\limits_{n=0}^{\infty}e^{-\pi\left(n+\frac{1}{2}\right)^2(r\pi(\lambda/d)^2)/2 } & \hspace{1mm} \\
                -\frac{1}{2\pi \beta\lambda^2}\sum\limits_{r=1}^{\infty}\frac{e^{\beta r \mu}}{r^2}\sum\limits_{n=0}^{\infty}e^{-\pi\left(n+\frac{1}{2}\right)^2(r\pi(2\lambda/d)^2)/2 } & \hspace{5mm}  \\
              \end{array} \right.
\end{align}
The 
following identities can be established from Poisson summation formula\cite{4} (see appendix),
\begin{equation}
    \sum_{n=0}^{\infty}e^{-\pi\left(n+\frac{1}{2}\right)^2 a }=\frac{1}{2\sqrt{a}}+\sum_{n=1}^{\infty}(-1)^n e^{-\pi n^2/a}, \quad a>0
\end{equation} and 
\begin{equation}
    \sum_{n=-\infty}^{\infty}e^{-\pi\left(n+\frac{1}{2}\right)^2 a }=\frac{1}{\sqrt{a}}+2\sum_{n=1}^{\infty}(-1)^n e^{-\pi n^2/a}, \quad a>0
\end{equation}\\
So, $\Phi_d(T,\mu)$ can therefore be expressed as follows,
\begin{eqnarray}
    \Phi_d(T,\mu)=\Phi_{bulk}(T,\mu) + \Phi_{Cas}(T,\mu)
\end{eqnarray}
In both cases the bulk contribution to grand potential is,
\begin{eqnarray}
\Phi_{bulk}=-\frac{d}{\beta(\sqrt{2\pi}\lambda)^3}\sum_{r=1}^{\infty}\frac{e^{\beta r \mu}}{r^{5/2}}
\end{eqnarray}
whereas, the second term is different for cases,
\begin{align}
\Phi_{Cas}(T,\mu) =-q\times \left\{ \begin{array}{cc} \sum\limits_{n=1}^{\infty}\sum\limits_{r=1}^{\infty}\frac{e^{\beta r \mu}}{r^{5/2}}(-1)^ne^{-2(nd/\lambda)^2/r}  \hspace{7mm}(Z)\\
     \sum\limits_{n=1}^{\infty}\sum\limits_{r=1}^{\infty}\frac{e^{\beta r \mu}}{r^{5/2}}(-1)^ne^{-2(nd/2\lambda)^2/r} \hspace{7mm}(A)\\
              \end{array} \right.
\end{align}
where $q=\frac{2d}{\beta(\sqrt{2\pi}\lambda)^3}$.
Note that, the surface term is absent for both  cases just like periodic boundary condition\cite{4}. Nevertheless, in any general system
the bulk as well the surface term,
 do not contribute to the
Casimir force, because the force due to the bulk term
 is counterbalanced by the same contribution
acting from outside the slabs when they are immersed in
the critical medium\cite{4,6} and the surface term 
does not change with the change of thickness of the slab.
Now the Casimir force can be obtained from Casimir potential 
through
 $-\partial_d \Phi_{Cas}^{(Z)}$. Now let us consider two separate cases: condensed and noncondensed case subject to Zaremba boundary condition-\\\\
In condensed phase $T\leq T_c$ $(\mu=0)$, BEC occurs in Bose gas and $\mu=0$. We write eq. (9) as
\begin{eqnarray}
    \Phi_{Cas}^{(Z)}(T,0)=-\frac{2}{\beta(\sqrt{2\pi})^3d^2}\sum_{n=1}^{\infty}(-1)^n(\lambda/d)^2     \sum_{r=1}^{\infty}\psi_n
    \big( (\lambda/d)^2r\big)
\end{eqnarray}
where we have defined $\psi_n(x)=\frac{e^{-2n^2/x}}{x^{5/2}}$. We now use the fact. $\lambda/d << 1$ and  the sum $\sum_{r=1}^{\infty}$ can be converted into an integral. Therefore we obtain,
\begin{eqnarray}
    \Phi_{Cas}^{(Z)}(T,0)=-\frac{1}{8\pi\beta d^2} \sum_{n=1}^{\infty}\frac{(-1)^n}{n^3}
     =\frac{\eta(3)}{8\pi\beta d^2}.
\end{eqnarray}
Where $\eta(s)=\sum_{n=0}^{\infty}\frac{(-1)^{n-1}}{n^s}$ is the Dirichlet eta function. Now, using the relation, $\eta(3)=\frac{3}{4}\zeta(3)$, where $\zeta(s)=\sum\limits_{n=1}^{\infty}n^{-s}$ is the Riemann zeta function. We finally obtain
\begin{equation}
     \Phi_{Cas}^{(Z)}(T,0)=\frac{3\zeta(3)}{32\pi}\frac{k_BT}{d^2}
\end{equation}
Note that the sign of $\Phi_{Cas}^{(Z)}(T,0)$ is positive in contrast to the situation described in references \cite{4}. Finally we have the Casimir force per unit area:
\begin{equation}
    F_c=-\partial_d \Phi_{Cas}^{(Z)}= \frac{3\zeta(3)}{16\pi}\frac{k_BT}{d^3}.
\end{equation}
which is repulsive.\\ \\
 Finally we consider the non-condensed phase
 with $T>T_c$ $(\mu<0)$.  The double sums in eq. (6) for $d>>\lambda$ can be estimated following Ref\cite{4}
\begin{eqnarray}
    \sum_{n=1}^{\infty}\sum_{r=1}^{\infty}\frac{e^{\beta r \mu}}{r^{5/2}}(-1)^ne^{-2(nd/\lambda)^2/r} \leq
    \frac{\zeta(5/2)}{e^{\sqrt{-8\beta\mu}d/\lambda}+1}
    =-O\left(e^{-\sqrt{-8\beta\mu}d/\lambda}\right).
\end{eqnarray}
As a result, the leading contributing to Casimir potential (eq 6) is decaying  exponentially just like Periodic, Dirichlet or Neumann boundary condition\cite{4}, but the force is positive unlike those boundary condition.
The Casimir force in noncondensed phase is therefore $F_C\propto exp (-d/\kappa^{(Z)}) $, where 
\begin{eqnarray}
\kappa^{(Z)} =\frac{\lambda}{4}\sqrt{\frac{2}{(-\mu)\beta}}
\end{eqnarray}
The Casimir decay length, $\kappa$ for mixed boundary condition is exactly equal to the Dirichlet/Neumann\cite{marek2} case\footnote{the definition of thermal wavelength with us and reference \cite{marek2} is different by a factor of 
$\frac{1}{\sqrt{2\pi}}$  }. Lets turn our attention towards the other case with anti-periodic boundary condition. Following the same procedure we find out the
Casimir force for $T<T_c$ is,
\begin{eqnarray}
    F_c^{(A)}=-\partial_d \Phi_{Cas}^{(A)}= \frac{3\zeta(3)}{2\pi}\frac{k_BT}{d^3}.
\end{eqnarray}
And in the non condensed phase the Casimir force is
$F_C ^{(A)}\propto exp(d/\kappa^{(A)})$ with 
\begin{eqnarray}
\kappa^{(A)} ={\lambda}\sqrt{\frac{1}{2(-\mu)\beta}}
\end{eqnarray}
\begin{center}
\begin{figure}
\begin{center}
\includegraphics[width=110mm,scale=0.5]{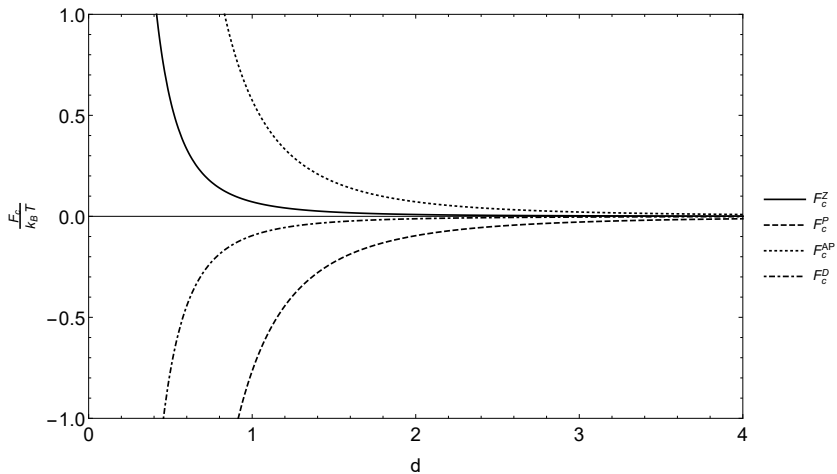}
\caption{The Casimir force of ideal Bose gas in condensed phase $(T<T_c)$.}
\label{fig.1}
\end{center}
\end{figure}
\end{center}
In  ref.\cite{marek2}, it has been established that decay length of Casimir force is related directly to the Bulk correlation length for Dirichlet (D), periodic (P) and Neumann (N) boundary conditions, $\frac{1}{2}\kappa^{(P)}=\kappa^{(D)}=\kappa^{(N)}=\xi=\frac{\lambda}{4}\sqrt{\frac{2}{(-\mu)\beta}} $. We have thus extended their relation in case of Zaremba and anti-periodic  boundary condition
$\frac{1}{2}\kappa^{(P)}=\kappa^{(D)}=\kappa^{(N)}=\kappa^{(Z)}=\frac{1}{2}
\kappa^{(AP)}$. But, most significant point of this calculation is the repulsive nature  of critical Casimir force,
both below and above critical temperature.
Nevertheless from here we can see that, like the other boundary conditions   upon approaching the phase containing the condensate
$(\mu \rightarrow 0)$ the range of force and the correlation length   diverges with the critical exponent $\nu = 1/2$ in Zaremba and anti-periodic boundary condition, identifying the characteristic nature of ideal gas.

\section{Concluding remarks}
In conclusion, 
we have identified the repulsive nature of Casimir force in both  phases of an ideal free Bose gas 
for thermal fluctuations in vacuum
subjected to Zaremba and anti-periodic boundary conditions  using the techniques of statistical mechanics. 
Extracting the final result in the current study is possible due to the identities of  eq. $(5)$ and $(6)$ (referred to as Jacobi identity in eq. (16) of ref.\cite{4}). If one compares these two identities 
with 
the corresponding  identities for Neumann, Dirichlet and periodic case, one can immediately notice the 
$(-1)^n$ in the second term with a sum over $n$ 
in eq (5) and (6).
This results in a  positive contribution to the
Casimir term in grand canonical
potential (eq. 12) unlike the cases with Neumann, Dirichlet or periodic boundary condition\cite{4,5}. The reason behind the $(-1)^n$ term in our identities of eq. $(5)$ and $(6)$ is the quantised momenta in eq. $(1)$, which is proportional to the half integers in this case whereas for Dirichlet, Neumann or periodic scenario quantised momenta whch are proportional to the integers.
As a consequence,
these boundary conditions 
result in a repulsive Casimir force.
In three dimensions, the magnitude of Casimir force with  Zaremba boundary condition is $\frac{3}{32}$ times the Casimir force in periodic boundary condition, while for antiperiodic scenario the Casimir force is attenuated by a factor of $\frac{3}{4}$ in comparison to the periodic case for any temperature\footnote{the Casimir force for Dirichlet/Neumann boundary condition are $\frac{1}{8}$ times than the periodic one\cite{6}}.
But these values will of course
change for
trapped bosonic systems\cite{6} which needs to be investigated as they are substantially
related with experimental detection of BEC\cite{exp,exp2}.
Point to note, this analysis is done solely for ideal gas without any sort of interaction, which proves that the repulsive behaviour is solely due to boundary condition. But at 
the same time the effect of interaction needs to be checked under these boundary conditions
to find out if any interesting scenario arises like ref \cite{marek5}. 
This is the next program that we wish to take up.
The current study unraveling
repulsive Casimir force in Bose Einstein condensate (BEC) could be a good prospect to different aspects of physics including nanotechnology\cite{app2,app3},  biology\cite{bio}, and complex networks\cite{net}. However, we should investigate the case of dynamical Casimir\cite{2,3,dyn}
effect in BEC, which 
has not yet been reported.
Such a study can not only  disentangle new features of BEC  but also shed new light on the relationship between Casimir force and the Cosmological constant, Dark energy and Dark matter\cite{dark,dark2}, especially in those models where scalar field BEC\cite{last1} and Axion BEC\cite{last2} are possible Dark matter candidates.
\\

\section{Appendix}
In this section we derive the mathematical identities described in eq. (5) and (6).
For appropriate functions $f$,
the Poisson summation formula can be stated as
\begin{equation}
    \sum_{n\in\mathbb{Z}} f(n) = \sum_{n\in\mathbb{Z}} \hat{f}(\nu)
\end{equation}
where, $\hat{f}(\nu)$ is the Fourier transform of $f(n)$.
\begin{equation}
    \hat{f}(\nu) = \int_{-\infty}^\infty dn\, f(n) e^{i 2\pi n\nu}
\end{equation}
Then,
\begin{align}
\sum_{n=-\infty}^{\infty} e^{-\pi a (n+1/2)^2} & =  \sum_{\nu=-\infty}^\infty \int_{-\infty}^\infty dn\, e^{-\pi a (n+1/2)^2 + i 2\pi n\nu} \nonumber\\
& =  \sum_{\nu=-\infty}^\infty \frac{1}{\sqrt{a}} e^{-i \nu\pi -\pi \nu^2/a} \nonumber\\
&= \sum_{n=-\infty}^\infty \frac{1}{\sqrt{a}} (-1)^n e^{-\pi n^2/a} \nonumber\\
&= \frac{1}{\sqrt{a}} + 2\sum_{n=1}^\infty \frac{1}{\sqrt{a}} (-1)^n e^{-\pi n^2/a} 
\end{align}
which is equation (6).
Now the left hand side of  eq. (20) 
can be re written as,

\begin{eqnarray}
\sum_{n=-\infty}^{\infty} e^{-\pi a (n+1/2)^2} &=&
    \underbrace{e^{-\pi a/4} }_\text{$n=0$} + 
    \underbrace{e^{-\pi a(3/2)^2} }_\text{$n=1$} +      \underbrace{e^{-\pi a(5/2)^2 } }_\text{$n=2$} +     \underbrace{e^{-\pi a(7/2)^2} }_\text{$n=3$} +     \underbrace{e^{-\pi a(9/2)^2} }_\text{$n=4$} +   \nonumber......\\  
   &&      \underbrace{e^{-\pi a/4} }_\text{$n=-1$} + 
    \underbrace{e^{-\pi a(3/2)^2} }_\text{$n=-2$} +     \underbrace{e^{-\pi a(5/2)^2 } }_\text{$n=-3$} +     \underbrace{e^{-\pi a(7/2)^2} }_\text{$n=-4$} +     \underbrace{e^{-\pi a(9/2)^2} }_\text{$n=-5$} +    ...... 
\end{eqnarray}
Therefore one can see the $n=0$ term matches with $n=-1$,
$n=1$ term matches with $n=-2$,
$n=2$ term matches with $n=-3$,
$n=3$ term matches with $n=4$ and so on.
As a result the eq. (21) can be written as,
\begin{eqnarray}
\sum_{n=-\infty}^{\infty} e^{-\pi a (n+1/2)^2} &=&
    2(\underbrace{e^{-\pi a/4} }_\text{$n=0$} + 
    \underbrace{e^{-\pi a(3/2)^2} }_\text{$n=1$} +      \underbrace{e^{-\pi a(5/2)^2 } }_\text{$n=2$} +     \underbrace{e^{-\pi a(7/2)^2} }_\text{$n=3$} +     \underbrace{e^{-\pi a(9/2)^2} }_\text{$n=4$} +   \nonumber......)\nonumber\\  
     &=&
     2\sum_{n=0}^{\infty}e^{-\pi\left(n+\frac{1}{2}\right)^2 a }
     \end{eqnarray}
     Combining equation (22) and (20)
     we obtain,
     \begin{eqnarray}
         \sum_{n=0}^{\infty}e^{-\pi\left(n+\frac{1}{2}\right)^2 a }=\frac{1}{2\sqrt{a}}+\sum_{n=1}^{\infty}(-1)^n e^{-\pi n^2/a}\nonumber
     \end{eqnarray}
     which is eq. (5)


\begin{thebibliography}{0}

\bibitem{casimir}
Casimir H. B. G., Proc. K. Ned. Akad. Wet., 51 (1948)
793.

\bibitem{2}
  Milton K. A., The Casimir Effect (World Scientific) 2001

\bibitem{3}
M. Bordag, U.Mohideen, V.M. Mostepanenko, Physics Reports 353 (2001) 1–205.

\bibitem{4}
Martin P. A. and Zagrebnov V. A., Europhys. Lett.,
73 (2006) 15.
\bibitem{5}
Biswas S., Eur. Phys. J. D, 42 (2007) 109.; Biswas S., J. Phys. A: Math. Theor., 40 (2007) 9969
\bibitem{6}Tongling Lin, Guozhen Su, Qiuping A. Wang and Jincan Chen,
2012 EPL 98 40010.

\bibitem{marek1}Marek Napiórkowski and Jarosław Piasecki
Phys. Rev. A 95, 063627. 


\bibitem{marek2}
Marek Napiórkowski, Jarosław Piasecki,
J Stat Phys (2014) 156:1136–1145.



\bibitem{marek4}
Napiórkowski, M., Piasecki, J. Phys. Rev. E 84, 061105
(2011)
\bibitem{marek5}
P. Jakubczyk, M. Napiorkowski and T. Sek,
EPL, 113 (2016) 30006 
\bibitem{nano}
C. Genet, Lambrecht. Reynaud,
	Eur. Phys. J. Special Topics 160, 183 (2008);
	\bibitem{milton33} Kimball A Milton et al 2012 J. Phys. A: Math. Theor. 45 374006

\bibitem{app} Andrea Gambassi 2009 J. Phys.: Conf. Ser. 161 012037
\bibitem{app2}
O. Kenneth, I. Klich, A. Mann, and M. Revzen,
DOI: 10.1103/PhysRevLett.89.033001.
\bibitem{app3}
Sandra J. Veen, Oleg Antoniuk, Bart Weber, Marco A. C. Potenza, Stefano Mazzoni, Peter Schall, and Gerard H. Wegdam,
Phys. Rev. Lett. 109, 248302.
F. M. Serry, D. Walliser, and G. J. Maclay, J.
Microelectromech. Syst. 4, 193 (1995); H. B. Chan,
V. A. Aksyuk, R. N. Kleiman, D. J. Bishop, and F.
Capasso, Science 291, 1941 (2001).


\bibitem{string}
A. A. Saharian, S. Kotanjyan,
 Eur. Phys. J. C (2011) 71: 1765.

\bibitem{qft1}
Xiang-hua Zhai and Xin-zhou Li, Phys. Rev. D 76, 047704 (2007).

\bibitem{qft2}
M. Asorey, J.M. Munoz-Castaneda,
Nuclear Physics B 874  (2013) 852–876.

\bibitem{bio}K Bradonjić, J D Swain, A Widom and Y N Srivastava, J. Phys.: Conf. Ser. 161 012035.

\bibitem{cos}
Gaurang Mahajan, Sudipta Sarkar, T. Padmanabhan, Physics Letters B 641 (2006) 6–10;
I. Brevik, K. A. Milton, S. D. Odintsov, and K. E. Osetrin
Phys. Rev. D 62, 064005.

\bibitem{experiment}
Garcia R. and Chan M. H., Phys. Rev. Lett., 83 (1999)
1187.
Garcia R. and Chan M. H., Phys. Rev. Lett., 88 (2002)
086101.
Fukuto M., Yano Y. F. and Pershan P. S., Phys. Rev.
Lett., 94 (2005) 135702.
\bibitem{ex2}
Ganshin A., Scheidemantel S., Garcia R. and Chan
M. H., Phys. Rev. Lett., 97 (2006) 075301.
Hartlein C., Helden L., Gambassi A., Dietrich S.
and Bechinger C., Nature, 451 (2008) 172.
\bibitem{dyn}
C. M. Wilson, G. Johansson, A. Pourkabirian, M. Simoen, J. R. Johansson, T. Duty, F. Nori and P. Delsing
Nature 479, 376–379.
\bibitem{dark}Kimball A Milton 2004 J. Phys. A: Math. Gen. 37 R209.
\bibitem{dark2}
Marek Szydłowski
 and Włodzimierz Godłowski
, Int. J. Mod. Phys. D 17, 343 (2008). 
\bibitem{exp}
 C. C. Bradley, C. A. Sackett, J. J. Tollett and R. G. Hulet, Phys. Rev. Lett. 75, 1687, (1995).
\bibitem{exp2} M. H. Anderson, J. R. Esher, M. R. Mathews, C. E. Wieman and E. A. Cornell, Science 269, 195, (1995).
\bibitem{net}
 Dorogovtsev, S. N.; Mendes, J. F. F. (2001). Phys. Rev. E 63: 056125.
\bibitem{last1}
Saurya Das and Rajat K Bhaduri 2015 Class. Quantum Grav. 32 105003.


\bibitem{last2}
Ozgur Erken, Pierre Sikivie, Heywood Tam, Qiaoli Yang,
arXiv:1111.3976.

\end{thebibliography}
\end{document}